\documentclass[12pt,preprint]{aastex}
\begin{document}
\newcommand{\msun}{M_{\odot}}
\newcommand{\kms}{\, {\rm km\, s}^{-1}}
\newcommand{\cm}{\, {\rm cm}}
\newcommand{\gm}{\, {\rm g}}
\newcommand{\erg}{\, {\rm erg}}
\newcommand{\kpc}{\, {\rm kpc}}
\newcommand{\mpc}{\, {\rm Mpc}}
\newcommand{\seg}{\, {\rm s}}
\newcommand{\kev}{\, {\rm keV}}
\newcommand{\hz}{\, {\rm Hz}}
\newcommand{\sr}{\, {\rm sr}}
\newcommand{\etal}{et al.\ }
\newcommand{\yr}{\, {\rm yr}}
\newcommand{\eq}{eq.\ }
\newcommand{\lya}{Ly$\alpha$\ }
\newcommand{\hi}{\mbox{H\,{\scriptsize I}\ }}
\newcommand{\hii}{\mbox{H\,{\scriptsize II}\ }}
\newcommand{\hei}{\mbox{He\,{\scriptsize I}\ }}
\newcommand{\heii}{\mbox{He\,{\scriptsize II}\ }}
\newcommand{\cii}{\mbox{C\,{\scriptsize II}\ }}
\newcommand{\ciis}{\mbox{C\,{\scriptsize II}${}^{\ast}$ }}
\newcommand{\nhi}{N_{HI}}
\def\arcsec{''\hskip-3pt .}

\title{
On the Importance of Local Sources of Radiation in Cosmological
Absorption Systems}
\author{Jordi Miralda-Escud\'e}
\affil{The Ohio State University}
\email{jordi@astronomy.ohio-state.edu}

\begin{abstract}

  An upper limit to the importance of local sources of radiation
compared to the cosmic background in cosmological absorption systems is
derived, as a simple consequence of the conservation of surface
brightness. The limit depends only on the rate of incidence of the
absorbers and the mean free path of the radiation. It is found that,
on average, the ionizing radiation intensity from local sources in Lyman
limit systems at $z>2$ must be less than half of the intensity of the
cosmic background. In absorbers with column densities much lower than
Lyman limit systems, the local source contribution must be negligible.
The limit on the ratio of local source to background intensities is
then applied to the class of damped \lya absorption systems with
detectable excited \cii lines. A cooling rate of the gas in these
systems has been measured by Wolfe \etal, who assumed that the balancing
heating source is photoelectric heating on dust by light at $\sim 1500$
\AA\ . The intensity from local star formation at this wavelength in
this class of damped \lya systems is found to be at most $\sim 3$ times
the background intensity. If the heating source is indeed photoelectric
heating of dust, the background created by sources associated with
damped \lya systems can then be estimated from the average cooling rates
measured in the absorbers. Current results yield a background intensity
higher than previous estimates based on observed galaxy and quasar
luminosity functions, although with a large uncertainty. The possibility
of other sources of heating, such as shock-heating in a turbulent
medium, should be explored.

\end{abstract}


\section{Introduction}

  In the study of the ionization of intergalactic absorption systems
observed in the spectra of quasars or other bright optical sources,
the question has often been raised of whether the ionizing radiation
that illuminates the absorbing gas is dominated by the cosmic
background, or by a local source associated with the absorber
(usually a galaxy in a dark matter halo). This paper presents a
straightforward argument that yields an upper limit to the importance of
any local source, demonstrating that local sources can be at most of
comparable importance to the cosmic background in typical damped \lya
and Lyman limit systems, and must be negligible compared to the
background in any class of absorption systems that have a rate of
incidence much higher than that of Lyman limit systems. This upper
limit is derived in \S 2, and an application to the origin of the
heating source in damped \lya systems with detected excited \cii
where the gas cooling rate can be derived is discussed in \S 3.

\section{The maximum average contribution of local sources}

  Consider the set of absorption systems with an observed column
density $N > N_{min}$ of some absorbing atom, with an average rate of
incidence per unit redshift on random lines of sight $r(z)$, as a
function of redshift $z$. Assume that each such absorber is associated
with a halo containing a local source of radiation. In an absorber at
impact parameter $b$ from a source of luminosity $L$, the absorbing gas
is at a distance larger than $b$ from the source, so the average flux
$F_l$ contributed by the local source on the absorbing atoms must be
$F_l < L/(4\pi b^2)$. The average surface brightness $s_l$ contributed
by the local source over the entire sky is simply $s_l=F_l/(4\pi)$. We
define $\bar F_l$ and $\bar s_l$ as the average values of the local flux
and surface brightness over all the absorption systems, which will
generally have a range of associated source luminosities and impact
parameters.

  Any absorber will be illuminated as well by all other sources
contributing to the cosmic background. A source that is associated with
another absorber at distance $x$ from the first one produces an
average flux
\begin{equation}
\bar F= \bar F_l {b^2\over x^2}\, e^{-x/\lambda(z_a)} ~,
\end{equation}
where $\lambda(z_a)$ is the mean free path of photons at redshift
$z_a$, and we assume for simplicity that $x$ is small compared to the
cosmic horizon at $z_a$. We can now imagine this flux spread over the
solid angle $\Omega$ over which the absorber is observed with column
density above $N_{min}$. For a spherical absorber with column density
$N_{min}$ at impact parameter $b$, we have $\Omega=\pi (b^2/x^2)$ (for
$x \gg b$), so the mean surface brightness within $\Omega$ is
\begin{equation}
 \bar s= { \bar F\over \Omega } = {\bar F_l\over \pi}\,
e^{-x/\lambda(z_a)} ~.
\end{equation}
Hence, the surface brightness of the cosmic background contributed by
all the sources associated with the absorbers of rate of incidence
$r(z)$ is
\begin{equation}
 s_b= \int_{z_a}^\infty dz\, r(z) {\bar F_l\over \pi}\,
e^{-(z-z_a)/\zeta(z_a)} ~,
\end{equation}
where $\zeta(z_a) = H(z_a)\, (1+z_a)\, \lambda(z_a)/c$ is the redshift
interval corresponding to the mean free path $\lambda(z_a)$, and
$H(z_a)$ is the Hubble constant at $z_a$. The ratio of the background
surface brightness to the average local surface brightness is then
\begin{equation}
 {s_b \over \bar s_l} = 4 \int_{z_a}^\infty dz\, r(z)\,
e^{-(z-z_a)/\zeta(z_a)} ~.
\label{rbl}
\end{equation}
We note that $s_b$ is the intensity of the {\it fraction} of the
background radiation that is contributed by sources associated with
the set of absorbers being considered. In practice, most of the
background radiation may be contributed by other sources and may then
have an intensity much larger than $s_b$. For example, the background
may be produced predominantly by highly luminous quasars and galaxies,
whereas most hydrogen absorbers may arise in low-mass halos containing
galaxies of low luminosity.

  Equation (\ref{rbl}) is of course easily generalized to include
cosmological effects that become important when the universe is not
highly opaque:
\begin{equation}
 {s_b(\nu) \over \bar s_l(\nu)} = 4 \int_{z_a}^\infty dz\, r(z)\,
{\bar F_l[\nu (1+z)/(1+z_a), z] \over \bar F_l (\nu, z_a)} \,
\left( { 1+z_a \over 1+z } \right)^3\, e^{-\tau(z,z_a)} ~,
\label{rblc}
\end{equation}
where $\nu$ is the frequency, $\bar F_l(\nu, z)$ is the average local
flux per unit frequency at frequency $\nu$ and redshift $z$, and the
optical depth from $z$ to $z_a$ is
\begin{equation}
 \tau(z,z_a) = \int_{z_a}^\infty {dz \over 1+z} \,
{c \over H(z) \lambda(z)} ~.
\end{equation}

  We now apply equation (\ref{rbl}) to hydrogen \lya absorption systems,
and consider photons near the hydrogen Lyman limit (where the mean free
path is shortest). Assuming that the column density distribution of
absorbers is $f(\nhi)\, d\nhi = \nhi^{-1.5}\, d\nhi$ (Petitjean \etal
1993), the mean free path of photons at frequencies just above the Lyman
limit is $\lambda = \lambda_{LL}/\sqrt{\pi}$, where $\lambda_{LL}$ is
the mean separation between Lyman limit systems with $\nhi > 1.6\times
10^{17} \cm^{-2}$ (see eq.\ [3] in Miralda-Escud\'e 2003). Hence, for
Lyman limit systems themselves, their rate of incidence is $r_{LL}(z) =
[\sqrt{\pi} \zeta(z)]^{-1}$, and using eq.\ (\ref{rbl}), we find that the
ratio of the background flux to the local source flux is
\begin{equation}
 {s_b \over \bar s_{l,LL}} > 4 \int_{z_a}^\infty {dz \over
\sqrt{\pi} \zeta(z) } \, e^{-(z-z_a)/\zeta(z_a)} \simeq 2.3 ~,
\label{rblll}
\end{equation}
where we have neglected the variation of $\zeta$ with $z$. This
approximation is adequate if the universe is opaque to ionizing photons,
an assumption that we have already made by neglecting the cosmological
terms from equation [\ref{rblc}]). We have used the $>$ sign as a
reminder that the background may be contributed mostly by sources that
are not associated with typical Lyman limit systems, and could then be
much more intense. Uncertainties in the observed column density
distribution can make only small changes to the numerical value in
equation (\ref{rblll}). Therefore, we have shown that
{\it on average, the intensity due to local sources in Lyman limit
systems cannot be more than about half the intensity of the cosmic
background}. 

  The above statement is valid at all redshifts when the universe
is opaque, or $z\gtrsim 1.5$. At low redshifts, the universe becomes
transparent and Lyman limit systems are found only in a small fraction
of lines of sight, so local sources may then be more important.

  For lower column density systems, the importance of local sources
must be even smaller in proportion to the inverse of the rate of
incidence, according to eq.\ (\ref{rbl}). For example, for systems with
$\nhi > 10^{15} \cm{-2}$, and assuming as before that $r(z) \propto
\nhi^{-0.5}$, the flux from a local source can contribute at most
4\% of the background flux.

  In a recent paper, Schaye (2004) claimed that local sources could,
under some assumptions, be dominant or comparable to the background in
absorption systems with $\nhi > 10^{15} f^{-2}$ to $10^{16} f^{-2}
\cm^{-2}$, depending on the model used for the column density
distribution (see his Table 1), where $f$ is a parameter depending on
the escape fraction of ionizing photons from Lyman break galaxies
(the dependence proportional to $f^{-2}$ is valid only for $r(z)
\propto \nhi^{-0.5}$). Schaye (2004) concluded that, for $f$ not much
smaller than unity, this may compromise the validity of ionization
models for the metal lines in these absorbers (e.g., Steidel \& Sargent
1992; Boksenberg \etal 2003). We have shown here that, in any class
absorbers with a rate of incidence that is not substantially smaller
than that of Lyman limit systems at $z\gtrsim 1.5$, local sources
cannot possibly dominate above the background. Schaye (2004) computed
the flux from local sources using models and assumptions for the
observed luminosity functions of galaxies and escape fractions, and he
estimated the background intensity separately from measurements of the
\lya forest transmitted flux. In Schaye's models in which radiation from
local sources can be of greater importance compared to the background
than the result we obtain here, the galaxies he postulated as local
sources would produce a background of higher intensity than he assumed.

  The argument presented here is not modified if the absorbers consist
of clusters of clouds, because the radiation from a local source that is
either in the same cloud or the same cluster of clouds as the absorbing
gas is subject to the same limit of equation (\ref{rbl}).

\section{Implications for the local ultraviolet flux in damped \lya systems}

  The radiation flux present in absorption systems is also important for
the heating rate of dense, metal-enriched gas in damped \lya systems.
The photoelectric effect on dust grains produced by ambient ultraviolet
photons may be the primary heating source of some of this gas, and is
believed to be the main heating source for the diffuse interstellar
medium in the Milky Way disk (e.g., Watson 1972; Weingartner \& Draine
2001). Wolfe \etal (2003a) determined the cooling rate through the $158
\, \mu{\rm m}$ emission by \cii in a set of damped \lya systems, by
measuring the column density of the excited \cii (or \ciis ) state. They
then derived the intensity of ultraviolet radiation that would be
required to balance this cooling from the photoelectric heating effect
on dust grains (Wolfe \etal 2003b, 2004).

  Photoelectric heating is caused by radiation of wavelengths $\lesssim
2000 \, {\rm \AA }$. In practice, only photons with wavelengths longer
than $912$ \AA\ are important, because the cosmic background intensity
decreases by a large factor at shorter wavelengths (due to the Lyman
break feature of star-forming galaxies), and because any photons at
shorter wavelengths would be absorbed by hydrogen before reaching the
interior of a damped \lya system. The maximum redshift range over which
the average photon at wavelength $\sim 1500$ \AA\ has propagated is
$\Delta z/(1+z) = (1-912/1500) $. Taking the mean free path
$\zeta_{UV}(z)$ of these photons to be half the maximum distance
travelled before they lose the ability to produce the photoelectric
effect due to redshift, we obtain $\zeta (z)/(1+z) \sim 0.2$. With a
rate of incidence for damped systems of $r(z)\simeq 0.25 [(1+z)/4]$
(Storrie-Lombardi \& Wolfe 2000), we find from eq.\ (\ref{rbl}):
\begin{equation}
 {s_b \over \bar s_{l,d}} > 4 \int_{z_a}^\infty dz\, r(z)\,
e^{-(z-z_a)/\zeta(z_a)} \simeq 4 r(z_a) \zeta(z_a) \simeq 0.8
 \left( {1+z\over 4} \right)^2 ~.
\label{rbld}
\end{equation}
Hence, we see that the ultraviolet radiation from the cosmic background
cannot be much less than the radiation from local star formation.

  Wolfe \etal (2004) detected absorption by \ciis for about half
of the damped \lya systems in a randomly selected sample. They found
that if the cooling rate inferred for these systems is balanced by
photoelectric heating on dust grains, then the mean ambient radiation
intensity at $1500$ \AA\ in this class of damped \lya absorbers is in
the range $10^{-19}$ to $10^{-18} \erg\cm^{-2}\seg^{-1}\hz^{-1}
\sr^{-1}$. According to our formula (\ref{rbld}), if this radiation is
present in all damped systems, then the background radiation should be
fainter than the average local source by only a factor $\sim 0.8$ at
$z=3$. If the damped systems without detected \ciis are a different
class of absorbers with much weaker local sources of radiation, then the
rate of incidence of the damped systems with detected \ciis is reduced
by a half, so equation (\ref{rbld}) shows the background must have an
intensity of at least $0.4$ times the average local source intensity, at
the lowest. If we take a value of $3\times 10^{-19} \erg\cm^{-2}
\seg^{-1}\hz^{-1}\sr^{-1}$ for the average ambient intensity, we find
that the background intensity that is contributed by the same sources
associated with this population of damped \lya systems would have to be
at least $10^{-19} \erg\cm^{-2}\seg^{-1}\hz^{-1}\sr^{-1}$. This is
already higher than the background estimated by the Haardt \& Madau
(1996) model based on the observed galaxy population at high redshift,
which is $3\times 10^{-20} \erg\cm^{-2}\seg^{-1}\hz^{-1}\sr^{-1}$ at
$1500$ \AA (see Fig.\ 1 in Wolfe \etal 2004).

  To make this into a precise estimate of the ultraviolet background
at high redshift, the average heating rate in damped \lya systems
required by the \ciis observations needs to be more accurately
determined. A possible problem with this method is that other sources
of heating may be important in addition to photoelectric heating of
dust, for the typical conditions in damped \lya systems. A likely source
of heating may be shock-heating due to cloud collisions. The observed
multiple absorption components in the metal lines associated with
damped \lya systems (Prochaska \& Wolfe 1997, 1998) indicate a level of
turbulence much higher than that in the Milky Way disk, implying that
cloud collisions must occur frequently and yielding a minimum heating
rate (McDonald \& Miralda-Escud\'e 1999). For example, for the typical
velocity separations among the absorption components of $v\sim 50 \kms$,
and dynamical times $t_d\sim 10^8$ years, the characteristic rate per
hydrogen atom would be $m_H v^2/ t_d \sim 10^{-26} \erg\seg^{-1}$,
comparable to the values found by Wolfe \etal (2004). However, the
cooling inferred from the \ciis column densities must take place in
cold neutral gas, at $T< 1000$ K, and the energy from cloud collisions
at the high velocities mentioned above would probably be radiated mostly
from gas at higher temperature. It is not clear how much of the
turbulent energy could be transferred to cold gas and radiated there.
We note also that although Wolfe \etal (2004) mention the observed
correlation of their derived heating rate with metallicity as evidence
that heating is due to the photoelectric effect on dust, this
correlation could be related to other expected correlations of
metallicity with halo velocity dispersion and impact parameter (which
affect the cloud collision velocities and dynamical time).

\section{Conclusions}

  We have presented a general formula that gives an upper limit to the
possible contribution of local sources to the total radiation flux
illuminating gas in absorption systems, compared to the contribution
from the cosmic background radiation. This upper limit demonstrates that
absorption systems that are more abundant than Lyman limit systems
cannot be affected by radiation dominated by a local source, on
average. A useful application of this upper limit is found when it is
applied to a class of absorbers that are required to be irradiated by a
minimum flux level: the background intensity generated by all the
sources associated with these absorbers can then be estimated. In the
case of damped \lya systems with detected \ciis absorption lines, we
find that if their gas is in fact heated by photoelectric heating on
dust at the level that has been claimed by Wolfe \etal (2004), then the
cosmic ultraviolet background is likely dominated by the sources
associated with these damped \lya absorbers and is higher than the
previously estimated background from known galaxies and quasars.

\acknowledgements

  I thank Joop Schaye and Art Wolfe for useful discussions.

\end{document}